\documentclass{article}

\usepackage[margin=1in]{geometry}

\usepackage{hyperref}

\usepackage{amsmath}
\usepackage{xspace}
\usepackage{longtable}
\usepackage{graphicx}
\usepackage{epstopdf}
\usepackage{subfigure}

\newtheorem{lemma}{Lemma}

\newcommand{\LL}{L\xspace}
\newcommand{\GG}{G\xspace}

\newcommand{\GS}{\ensuremath{\mathcal{G}}\xspace}
\newcommand{\OS}{\ensuremath{\mathcal{O}}\xspace}

\title{Combinatorial algorithm for counting small induced \\graphs and orbits}

\author{Toma\v{z} Ho\v{c}evar \\ University of Ljubljana \\ tomaz.hocevar@fri.uni-lj.si \and
Janez Dem\v{s}ar \\ University of Ljubljana \\ janez.demsar@fri.uni-lj.si}

\date{\vspace{1em}}

\begin{document}

\maketitle

\begin{abstract}
Graphlet analysis is an approach to network analysis that is particularly popular in bioinformatics. We show how to set up a system of linear equations that relate the orbit counts and can be used in an algorithm that is significantly faster than the existing approaches based on direct enumeration of graphlets. The algorithm requires existence of a vertex with certain properties; we show that such vertex exists for graphlets of arbitrary size, except for complete graphs and $C_4$, which are treated separately. Empirical analysis of running time agrees with the theoretical results.
\end{abstract}


\section{Introduction}

Analysis of networks plays a prominent role in data mining, from learning patterns \cite{Kuramochi2001} and predicting new links in social networks \cite{Backstrom2011,Liben-Nowell2007}, to inferring gene functions from protein-protein interaction networks \cite{Milenkovic2008_function} in bioinformatics. Many methods rely on the concept of node similarity, which is typically defined in a local sense, {\em e.g.} two nodes are similar if they share a large number of neighbours. Such definitions are insufficient for detecting the role of the node. A typical social structure includes hubs, followers, adversaries and intermediaries between groups. While local similarity definitions treat the hub and its adjacent nodes as similar, a role-based similarity would consider the hubs as similar disregarding their distance in the graph.

A popular approach in bioinformatics extracts the node's local topology by counting the small connected induced subgraphs (called {\em graphlets})~\cite{Przulj2004_graphlets}, which the node touches, and, when a more detailed picture is required, the node's position ({\em orbit}) \cite{Przulj2007_orbits} in those graphs.

\begin{figure}[hbt]
\begin{center}
\includegraphics[scale=0.7]{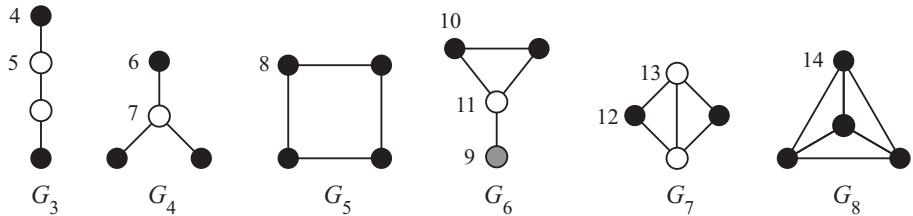}
\caption{Four-node graphlets ($\GS_4$). Vertices marked by the same color belong to the same orbit within a graphlet.}
\label{four-node}
\end{center}
\end{figure}

Let $\GS_k$ be a set of all non-isomorphic connected simple graphs on $k$ nodes, and let $\GG\in\GS_k$. The orbit of a vertex $v\in\GG$ is a set of all vertices $a(v)$, $a\in Aut(\GG)$. Let $\OS_k$ be a set of orbits for all $v\in\GG$ and for all $\GG\in \GS_k$. Pr\v zulj~\cite{Przulj2007_orbits} numbered the 30 graphs in $\GS_2$, $\GS_3$, $\GS_4$ and $\GS_5$ and the corresponding 73 orbits; we will use her enumeration in the examples in this paper. Figure \ref{four-node} illustrates all four-node graphlets and orbits of their nodes.

Let $H=(V, E)$ be the host graph and let $x\in H$. Vertex $x$ participates in a number of subgraphs $\GG\in\GS_k$ induced in $H$, in which it appears in different orbits $O_j\in\OS_k$. Let $o_j$ be the number of times $x$ appears in orbit $O_j$ in induced subgraphs from $\GS_k$.

\begin{figure}[tbp]
\begin{center}
\includegraphics[scale=0.70]{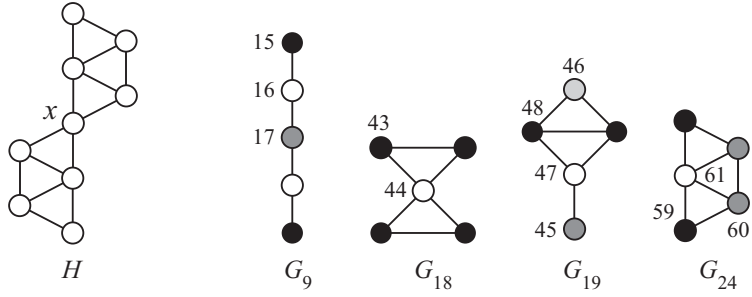}
\caption{A host graph $H$ and graphs $G_{9}$, $G_{18}$, $G_{19}$ and $G_{24}$ from $\GS_{5}$. Graphs and orbits are numbered as in~\cite{Przulj2007_orbits}.}
\label{f-example-counts}
\end{center}
\end{figure}

An example is shown in Figure~\ref{f-example-counts}. The orbit count $o_{17}$ of vertex $x$ is 9 since $x$ appears in nine paths $G_9$ as the central vertex (note that the paths must be induced). Other orbit counts for $G_9$, $o_{15}$ and $o_{16}$, are 0 and 4, respectively: $x$ does not appear as the end vertex ($O_{15})$ of $G_9$ in $H$, but it appears four times in the role of the node between the center and the end ($O_{16}$). For a few more examples, $o_{44}=1$, $o_{47}=4$, and $o_{59}=2$; all other orbit counts of 4-node graphlets are 0.

The {\em orbit count distribution} is a $|\OS_k|$-dimensional vector of $o_j$ for all $O_j\in\OS_k$. The orbit count distribution represents a signature of the node: it contains a description of the node's neighborhood and the node's position (``role'') within it. As such, this distribution is a useful feature vector for various network analysis tasks.

We will describe an algorithm for computation of orbit count distributions for all vertices $x\in V$ for subgraphs of size $k$ and show -- both theoretically as well as empirically -- that its time complexity on the real-world sparse graphs is lower by an order of magnitude in comparison with enumeration-based approaches.

\subsection{Preliminaries}

We will use the following notation.

\begin{longtable}{rp{0.6\linewidth}}
$H=(V, E)$ & host graph within which we count the graphlets and orbits\\
$n$ & number of nodes of $H$; $n=|V|$\\
$e$ & number of $H$'s edges; $e=|E|$\\
$d(v)$ & degree of node $v$\\
$d$ & maximal node degree in $H$; $d=\max_{v\in V} d(v)$\\
$N(v)$ & set of neighbours of vertex $v\in V$\\
$N(v_1, v_2, \ldots, v_j)$ & set of common neighbours of $v_1$, $v_2$,\ldots, $v_j$;\newline $N(v_1, v_2, \ldots, v_j) = N(v_1)\cap N(v_2)\cap\ldots\cap N(v_j)$\\
$N(\mathcal{S})$ & common neighbours of nodes in the set $\mathcal{S}\subset{V}$;\newline $N(\mathcal{S}) = \cap_{v\in\mathcal{S}} N(v)$\\
$c(v)$, $c(v_1, v_2, \ldots, v_j)$, $c(\mathcal{S})$ & number of common neighbours of vertex $v$, of vertices $v_1, v_2, \ldots, v_j$, and of vertices from set $\mathcal{S}$, respectively; that is, $c(v) = |N(v)|$, $c(v_1, v_2, \ldots, v_j) = |N(v_1, v_2, \ldots, v_j)|$, $c(\mathcal{S}) = |N(\mathcal{S})|$\\
$\mathcal{G}_k$ & set of all graphlets with $k$ nodes\\
$G_i$ & graphlet $i$, according to some enumeration\\
$O_j$ & orbit $j$, according to some enumeration\\
$o_j(v), o_j$ & the number of times the node $v$ appears in an induced subgraph in orbit $j$; since $v$ will be obvious, we will use the shorter notation $o_j$\\
$m(i)$ & index of the graphlet containing the orbit $O_i$, {\em e.g.} $m(16)=9$\\
\end{longtable}

Let $K=(V_K, E_K)$ be a subgraph of $J=(V_J, E_J)$, and let $v\in V_K$. We will denote $J$'s vertex that corresponds to $v$ by $v^J$. If there are multiple isomorphic embeddings of $K$ in $J$, $v^J$ refers to one of them. Similarly, if $S\subseteq V_K$, then the corresponding vertices in $J$ are denoted by $S^J$.

\subsection{Related work}

The most basic case of counting induced patterns in graphs is that of counting triangles. Itai and Rodeh~\cite{Itai1978_triangles} showed that this can be done faster than by exhaustive enumeration in $O(n^3)$ time. Raising the graph's adjacency matrix $A$ to the third power gives the number of paths of length 3 between pairs of nodes. Element $A^3_{x,x}$ represents the number of paths of length 3 that start and finish in the node $x$, which corresponds to the number of triangles that include $x$. The total number of triangles is then $\frac{1}{6} \sum_{x \in G} A^3_{x,x}$. Note that the same triangle is counted twice for each of its three nodes. The time complexity of this procedure equals that of multiplying two matrices, which is faster than exhaustive enumeration of triangles in dense graphs. A natural extension of this result is to larger cliques. Nesetril and Poljak~\cite{Nesetril1985_subgraph} studied the problem of detecting a clique of size $k$ in a graph with $n$ nodes. They showed that this problem can be solved faster than with the straight-forward $O(n^k)$ solution. Their approach reduces the original problem to detection of triangles in a graph with $O(n^{k/3})$ nodes. Since we can detect triangles faster than in $O(n^3)$ with fast matrix multiplication algorithms, we can also detect cliques of size $k$ faster than $O(n^k)$.

Counting all non-induced subgraphs is as hard as counting all induced subgraphs because they are connected through a system of linear equations. Despite this it is sometimes beneficial to compute induced counts from non-induced ones. Rapid Graphlet Enumerator (RAGE)~\cite{Marcus2012_rage} takes this approach for counting four-node graphlets. Instead of counting induced subgraphs directly, it reconstructs them from counts of non-induced subgraphs. For computing the latter, it uses specifically crafted methods for each of the 6 possible subgraphs ($P_4$, claw, $C_4$, paw, diamond and $K_4$). The time complexity of counting non-induced cycles and complete graphs is $O(e \cdot d + e^2)$, while counting other subgraphs runs in $O(e \cdot d)$. However, the run-time of counting cycles and cliques in real-world networks is usually much lower.

Some approaches exploit the relations between the numbers of occurrences of induced subgraphs in a graph. Kloks~{\em et al}.~\cite{Kloks2000_eq} showed how to construct a system of equations that allows computing the number of occurrences of all six possible induced four-node subgraphs if we know the count of any of them. The time complexity of setting up the system equals the time complexity of multiplying two square matrices of size $n$. Kowaluk {\em et al.}~\cite{Kowaluk2013_equations} generalized the result by Kloks to counting subgraph patterns of arbitrary size. Their solution depends on the size of the independent set in the pattern graph and relies on fast matrix multiplication techniques. They also provide an analysis of their approach on sparse graphs, where they avoid matrix multiplications and derive the time bounds in terms of the number of edges in the graph.

Floderus~{\em et al}.~\cite{Floderus2012_easypatterns} researched whether some induced subgraphs are easier to count than others as is the case with non-induced subgraphs. For example, we can count non-induced stars with $k$ nodes, $\sum_{x \in V} \binom{c(x)}{k-1}$, in linear time. They conjectured that all induced subgraphs are equally hard to count. They showed that the time complexity in terms of the size of G for counting any pattern graph $H$ on $k$ nodes in graph $G$ is at least as high as counting independent sets on $k$ nodes in terms of the size of $G$.

Vassilevska and Williams~\cite{Vassilevska2009} studied the problem of finding and counting individual non-induced subgraphs. Their results depend on the size $s$ of the independent set in the pattern graph and rely on efficient computations of matrix permanents and not on fast matrix multiplication techniques like some other approaches. If we restrict the problem to counting small patterns and therefore treat $k$ and $s$ as small constants, their approach counts a non-induced pattern in $O(n^{k-s+2})$ time. This is an improvement over a simple enumeration when $s \geq 3$. Kowaluk~{\em et al.}~\cite{Kowaluk2013_equations} also improved on the result of Vassilevska and Williams when $s=2$. Alon et al.~\cite{Alon1997} developed algorithms for counting non-induced cycles with 3 to 7 nodes in $O(n^\omega)$, where $\omega$ represents the exponent of matrix multiplication algorithms. 

Alon~{\em et al.}~\cite{Alon1995_color-coding} introduced the color-coding technique for finding simple paths and cycles in graphs. Their technique is applicable not just to paths and cycles but also to other patterns with a small treewidth. The authors of \cite{Alon2008} used such color-coding approach to approximate a `treelet' distribution (frequency of non-induced trees) for trees with up to 10 nodes.

\subsection{Outline of the proposed algorithm}

We will derive a system of linear equations that relate the orbit counts of a fixed node for graphlets with $k$ vertices, like equation (\ref{eq-example-59}) in the example below. The coefficients on the left-hand sides reflect the symmetries in the graphlets and do not depend on the host graph, so they are derived in advance. The right-hand sides are computed as sums over graphlets with $k-1$ vertices induced in the host graph $H$, and the sums include terms that represent the number of common neighbours of certain vertices in the embeddings of graphlets in $H$.

The resulting system of equations will be triangular and have a rank of $|\mathcal{O}_k| - 1$. We can efficiently enumerate the complete graphlet, after which the system of equations for the remaining orbit counts can be solved using integer arithmetic, thus avoiding any numerical errors.

\subsection{Original contributions}


We already presented the original idea of the algorithm in a recent article in Bioinformatics~\cite{Hocevar2014}, in which we focused on its use in genetics and avoided formal descriptions and analysis. In this paper we
\begin{enumerate}
\item present the algorithm more formally;
\item describe a general method for derivation of the system of equations relating the orbit counts (Section~\ref{s-general-relations});
\item generalize it to induced subgraphs of arbitrary size; in particular, we prove that the system of equations with the properties required for the efficient implementation of the algorithm exists for any $k\ge 4$ (Section~\ref{sec:y});
\item provide worst time-complexity analysis (Section~\ref{sec:worst}) and the analysis of the expected time complexity on random graphs (Section~\ref{sec:expected});
\item empirically explore the efficiency of the orbit counting algorithm and compare it with the theoretical results (Section~\ref{sec:empirical});
\end{enumerate}

The remainder of the paper is composed of two parts. In the next section we show a technique for building the system of equations with desired properties, and in the following section we present an algorithm based on them and analyze its time- and space-complexity.

\section{Relations between orbit counts\label{s-relations-counts}}

We will show how to construct linear relations between a chosen orbit count $o_j$ and some orbits belonging to graphlets with a larger number of edges than the one corresponding to $o_j$. We will first provide an example for $o_{59}$, and then present a general derivation.

\subsection{Example of derivation}

Orbit counts $o_{59}$, $o_{65}$, $o_{68}$ and $o_{70}$ are related as follows.

\begin{equation} 
o_{59} + 4o_{65} + 2o_{68} + 6o_{70} = \sum_{\substack{
x_1,x_2,x_3: \\
x_1 < x_2 \wedge x_3 \notin N(x), \\ 
H[\{x,x_1,x_2,x_3\}] \cong G_7}} 
\left[\left(c(x_1,x_3)- 1\right) + \left(c(x_2,x_3) - 1\right)\right].
\label{eq-example-59}
\end{equation}

\begin{figure}
\begin{center}
\subfigure[$G_7$]{
\includegraphics[scale=0.65]{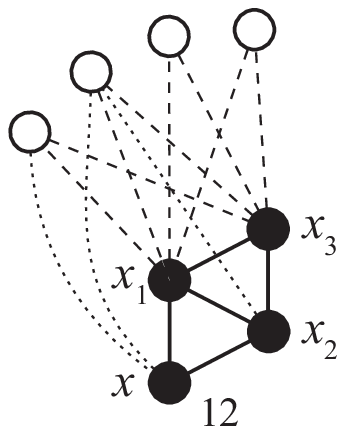}
\label{f-relation-y}}
~~
\subfigure[$G_{24}$]{
\includegraphics[scale=0.65]{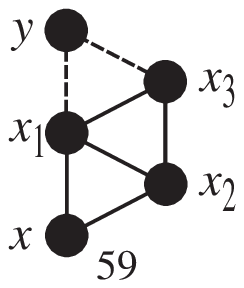}
\label{f-relation-24}}
~~
\subfigure[$G_{26}$]{
\includegraphics[scale=0.65]{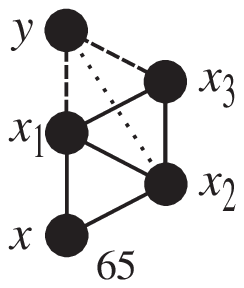}
\label{f-relation-26}}
~~
\subfigure[$G_{27}$]{
\includegraphics[scale=0.65]{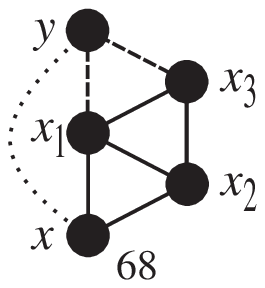}
\label{f-relation-27}}
~~
\subfigure[$G_{28}$]{
\includegraphics[scale=0.65]{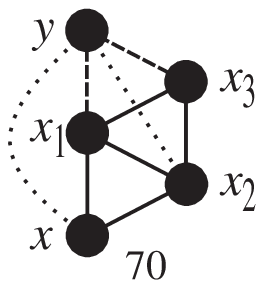}
\label{f-relation-28}}
\end{center}
\label{f-relation}
\caption{Example for constructing the relation between $o_{59}$, $o_{65}$, $o_{68}$, and $o_{70}$. Dashed lines represent edges due to condition $y\in N(x_1, x_3)$, and dotted lines represent edges that make the resulting induced graph isomorphic to $G_{24}$, $G_{26}$, $G_{68}$, or $G_{70}$, thus putting $x$ into one of the above orbits.}
\end{figure}

Let a fixed vertex $x$, for which we will compute the count $o_{59}$ from $G_{24}$, appear in orbit $o_{12}$ of an induced subgraph $G_7$ within the host $H$ (Figure~\ref{f-relation-y}). We assign labels $x_1$, $x_2$ and $x_3$ to the other three vertices as in Figure~\ref{f-relation-y}. 

$G_{24}$ and $G_7$ differ by a vertex adjacent to the vertices labelled $x_1$ and $x_3$. Let us thus observe the common neighbours $y$ of nodes $x_1$ and $x_3$, that is, $y\in N(x_1, x_3)$. Besides having an edge with $x_1$ and $x_3$, vertices in $N(x_1, x_3)$ can also be adjacent to $x_2$ and/or $x$, resulting in four possible induced five-node graphs on $\{x, x_1, x_2, x_3, y\}$, which are shown in Figures~\ref{f-relation-24}--\ref{f-relation-28}. Therefore, $o'_{59} + o'_{65} + o'_{68} + o'_{70} = |N(x_1, x_3)| - 1 = c(x_1, x_3) - 1$, where $o'_j$ represents the orbit counts for $x$ with this particular selection and marking of vertices $x_1$, $x_2$ and $x_3$. The term $-1$ is needed since $x_2$ is already one of the neighbours of $x_1$ and $x_3$.

We get (\ref{eq-example-59}) by summing this equation over all induced graphs $G_7$ with $x$ in orbit $12$. We count both, $y\in N(x_1, x_3)$ and $y\in N(x_2, x_3)$ at the same time because we can obtain $G_{24}$ by attaching $y$ to either $(x_1,x_3)$ or $(x_2,x_3)$. Conditions $x_1 < x_2$ and $x_3 \notin N(x)$ guarantee that we enumerate every $G_7$ with node $x$ in orbit $O_{12}$ exactly once.

To derive the coefficients on the left side of (\ref{eq-example-59}), let us first consider $O_{65}$. The sum on the right side of~(\ref{eq-example-59}) will count the graphlet in Figure~\ref{f-relation-26} four-times: (i) like shown in Fig.~\ref{f-relation-26}, that is, $y\in N(x_1, x_3)$, (ii) labels as shown in the figure, but with $y\in N(x_2, x_3)$ (the same picture except that the edge $(y, x_1)$ is dotted instead of $(y, x_2)$), (iii) and (iv) like the first two configurations except with switched labels for $y$ and $x_3$.

Orbit $O_{68}$ (Fig.~\ref{f-relation-27}) is counted twice due to switching of $y$ and $x_2$.

Orbit $O_{70}$ (Fig.~\ref{f-relation-28}) is counted six times. First, $y$ can be swapped with $x_1$ and $x_2$. In each of those three configurations, $y$ appears both in $N(x_1, x_3)$ and $N(x_2, x_3)$, which gives a total of six combinations.

Orbit $O_{59}$ (Fig.~\ref{f-relation-24}) is counted once. Node $y$ will be in exactly one of the sets $N(x_1, x_3)$, $N(x_2, x_3)$ and no other subset of nodes induces $G_7$ with $x$ in $O_{12}$.

Coefficients on the left-hand side thus reflect the number of times that each corresponding orbit will be overcounted by the sums on the right-hand side.

\subsection{Derivation of general relations between orbit counts\label{s-general-relations}}

\begin{figure}
\begin{center}
\includegraphics[scale=0.7]{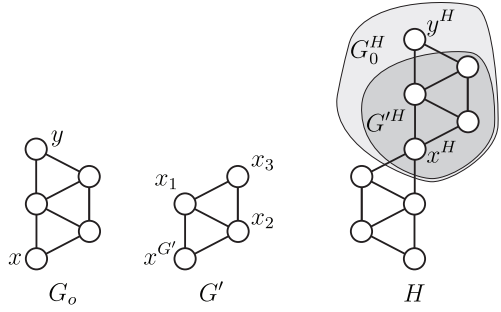}
\end{center}
\caption{Illustration of notation. $G_o$ is the subgraph with the node $x$ in the observed orbit and $G' = G\setminus\{y\}$. The embeddings of $G_o$ and $G'$ in $H$ are referred to as $G_o^H$ and $G'^H$, respectively. The vertices corresponding to $x$ and $y$ are marked by $x^H$ and $y^H$, analogously, the other three vertices from $G'$ are labelled by $x_1^H$,  $x_2^H$ and $x_3^H$ (omitted for clarity).}
\label{fig-notation-iso}
\end{figure}

We will present a general procedure for deriving similar relations for an arbitrary orbit $O_j$ in a connected simple $k$-node graphlet $G_o=(V_G, E_G)$ ($o=m(j)$). We denote the $G_o$'s node that is in orbit $O_j$ by $x$; if there are multiple such nodes, we pick one. Next, we choose a node $y\neq x$, such that $G'=G_o\setminus\{y\}$ is still a connected graph.\footnote{The condition that $G'$ should be connected suffices for derivation of relations in this section. We will impose additional constraints on $y$ later in order to ensure an efficient implementation of the algorithm.} $G'$ is a $(k-1)$-node graph; $G'=(V_{G'}, E_{G'})$, where $V_{G'}=V_G\setminus\{y\}$. According to our notation, $x^{G'}$ is the node in $G'$ that corresponds to $x$ in $G_o$; let $O_m$ be its orbit. We label the remaining $k-2$ nodes with $x_1, x_2, ..., x_{k-2}$. Notation is illustrated in Fig.~\ref{fig-notation-iso}. 

For example, when considering the orbit $O_{59}$ from $G_{24}$, we removed the node $y$ and got $G_7$, with $x^{G'}$ in orbit $O_{12}$.

We now consider the possible extensions of $G'$ to $G_o$. Let $E\subset V_{G'}$ be a set of nodes such that adding a new vertex $y$ connected to all vertices in $E$ yields $G_o$ with $x$ in orbit $O_j$. Let $\mathcal{E}$ be a set of all such subsets $E$. In the introductory example, $G_7$ can be extended to $G_{24}$ by attaching $y$ to either $x_1$  and $x_3$ or to $x_2$ and $x_3$, hence $\mathcal{E}=\{\{x_1, x_3\}, \{x_2, x_3\}\}$. 

Let $G'^H$ be some particular occurrence of $G'$ in $H$. To count $o_j$ for the node $x^H$ (the node in $H$ to which $x$ maps), we need to explore the extensions of $G'^H$ to $G_o^H$.

A necessary (but insufficient) condition to put $x^H$ into $O_j$ is that the additional node $y$ is a common neighbour of all vertices $E^H$ for one of $E^H\in\mathcal{E^H}$ (with respect to the particular occurrence of $G'$ in $H$). There are at most
\begin{equation}
\sum_{E\in\mathcal{E}} \left(c(E^H) - c(E^{G'})\right)
\label{f-right-hand}
\end{equation}
candidate nodes $y$; $c(E^{G'})$ represents the number of neighbours of $E$ that are already in $G'$ ({\em i.e.} $x_i$) and cannot be mapped to $y$. The sum (\ref{f-right-hand}) represents the term in the sum in the right side of the relation. In the introductory example, we have $\mathcal{E}=\{\{x_1, x_3\}, \{x_2, x_3\}\}$ and $c(E^{G'})=1$; $x_1$ and $x_3$ (as well as $x_2$ and $x_3$) have one common neighbour in $G'$. Thus $\sum_{E\in\mathcal{E}} \left(c(E^H) - c(E^{G'})\right) = (c(x_1,x_3) - 1) + (c(x_2,x_3) - 1)$. The right side in (\ref{eq-example-59}) sums this over all unique occurrences of $G'$ with x in $O_{12}$ within $H$.

Condition $y\in N(E^H)$ (for some $E^H\in\mathcal{E}^{H}$) is not sufficient since it allows $y$ to be connected to some additional nodes in $G'^H$, which puts $x^H$ in different orbits. In the Fig.~\ref{f-relation} of the introductory example, the necessary edges are shown with dashed lines and the extra edges with dotted ones. The vertex $x$ is in $O_{59}$ if $y$ is not connected to any other nodes, or in $O_{65}$ if it is also connected to $x_2$, and so forth.

The counts for these orbits, $o_p$, represent the variables on the left side of the relation. Node $y$ can be connected to additional $k - 1 - |E|$ ($E\in\mathcal{E}$) nodes, hence the left side can have at most $2^{k - 1 - |E|}$ terms.

\begin{figure}[tb]
\begin{center}
\includegraphics[scale=0.7]{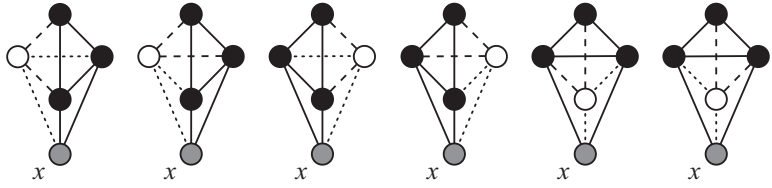}
\end{center}
\caption{Illustration of symmetries resulting in coefficient 6 for $o_{70}$ in (\ref{eq-example-59}).}
\label{f-coefficients}
\end{figure}

The corresponding coefficients at orbit $p$ represent the over-counts for these orbits, that is, the number of times that (\ref{f-right-hand}) counts the same occurrence of $G^* = G_{m(p)}$ (the graphlet containing the orbit $p$) within $H$. $G^*$ is obtained by extending $G'$ with $y\in N(E)$. The coefficient thus equals the number of ways in which $G'$ can be extended to $G^*$ with a fixed node $x$. Figure~\ref{f-coefficients} illustrates all 6 ways in which Equation \ref{eq-example-59} considers the same occurrence of node $x$ in orbit $O_{70}$. To compute the coefficient in general, we have to consider all induced occurrences of $G'$ in $G^*$ (with a fixed point $x$), which is the same as considering nodes $z \in V_{G^*}$ whose removal results in $G'$ with $x^{G'}$ in orbit $O_p$. For every such case we increase the coefficient by the number of extensions $E \in \mathcal{E}$ such that node $z$ is connected to the extension nodes, i.e. $N(z) \supseteq E^{G^*}$.

The general procedure for relating the orbit count $o_j$ with counts of orbits with higher indices like~(\ref{eq-example-59}) is thus as follows.
\begin{enumerate}
\item Let $G = G_{m(j)}$ be the graphlet that contains $O_j$.
\item Pick node $y$ such that $y \neq x$ and the $G' = G\setminus\{y\}$ is a connected graph.
\item The right side sums over all occurrences of $G'$; conditions are used to put $x$ into $O_j$ and to ensure that each occurrence of $G'$ is considered exactly once.
\item The terms within the sum are as given in~(\ref{f-right-hand})
\item The terms on the left side refer to orbits of $x$ within graphlet $G$ and graphlets with additional connections between $y$ and other nodes in $G$.
\item The coefficient for each term is independent of the host graph $H$ and is determined as explained.
\end{enumerate}

\subsection{Additional constraints on selection of \texorpdfstring{$y$}{y}}
\label{sec:y}

In the preceding derivation, the only limitation on selection of vertex $y$ was that the remaining graphlet is still connected. Different choices of $y$ yield different equations. With the coefficients independent of the host graph and known in advance, the time consuming part of using these equations to calculate orbit counts is the computation of the right-hand side terms. To speed it up, we impose some additional constraints on the choice of the node $y$: the restraints will be such that the right-hand sides will contain only the counts $c(S)$ in which either $|S|<k-2$, or equal $|S|=k-2$ with the nodes in $S$ forming a connected subgraph of $G_k$. This will allow pre-calculation and caching of all $c(S)$ needed for computation of right-hand sides.

For efficient precomputation, vertex $y\ne x$ must meet the following criteria:
\begin{enumerate}
\item $d(y) \leq k-2$,
\item $\GG\setminus \{y\}$ is a connected graph,
\item if $d(y) = k-2$, the neighbours of $y$ induce a connected graph,
\end{enumerate}
where $d(y)$ represents the degree of $y$.

We will prove that such a vertex exists in any graphlet $k \geq 4$ and all possible $x$, except for complete graphlets (all vertices violate the first condition) and for the cycle on four points, $C_4$ (all vertices violate the last condition).

\begin{figure}
\begin{center}
\begin{tabular}{c@{\hspace{7mm}}c@{\hspace{7mm}}c}
\subfigure[$u=1$]{
\includegraphics[width=0.25\textwidth]{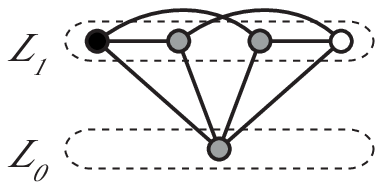}
\label{f-case-1}}
&
\subfigure[$u=2  \wedge |\LL_2| > 1$]{
\includegraphics[width=0.25\textwidth]{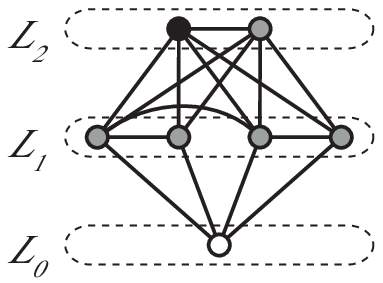}
\label{f-case-2}}
&
\subfigure[$u=2 \wedge  |\LL_2| = 1 \wedge  k = 4$]{
\includegraphics[width=0.25\textwidth]{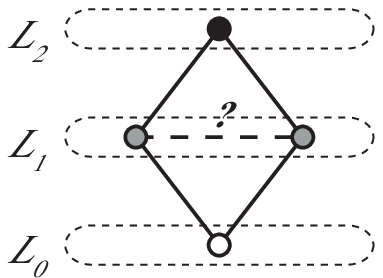}
\label{f-case-3}}
\end{tabular}
\\[0.5cm]
\begin{tabular}{c@{\hspace{7mm}}c@{\hspace{7mm}}c}
\subfigure[$u=2 \wedge |\LL_2| = 1 \wedge k~\ge~5 \wedge d(l_1)=k-1$]{
\includegraphics[width=0.25\textwidth]{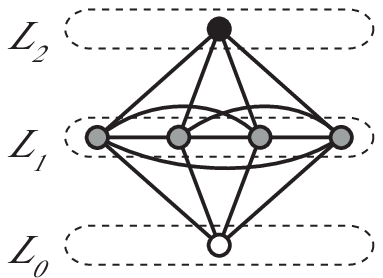}
\label{f-case-4}}
&
\subfigure[$u=2 \wedge |\LL_2| = 1 \wedge k~\ge~5 \wedge d(l_1) < k-2$]{
\includegraphics[width=0.25\textwidth]{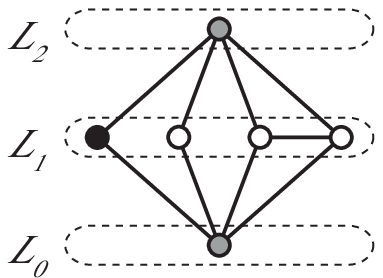}
\label{f-case-6}}
&
\subfigure[$u=2 \wedge |\LL_2| = 1 \wedge k~\ge~5 \wedge d(l_1) = k-2$]{
\includegraphics[width=0.25\textwidth]{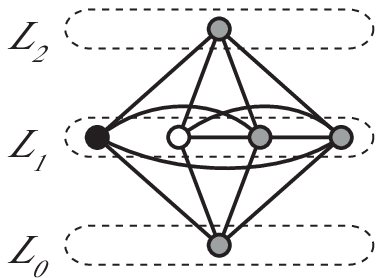}
\label{f-case-5}}
\end{tabular}
\end{center}
\caption{Illustrations of different cases. Node at the bottom level $\LL_0$  represents node $x$. The selected node $y$ is colored black and its neighbours are indicated with a gray color. }
\label{f-cases}
\end{figure}

Let $\LL_i$ represent the set of vertices at a distance $i$ from $x$ (see Fig.~\ref{f-cases}). Let $l_i$ be the vertex in $\LL_i$ with the smallest degree. Let $\LL_u$ be the last non-empty set, and, accordingly, $l_u$ the vertex with the smallest degree among the vertices farthest from $x$. We will show that $l_u$ fulfils the conditions in most cases, except in some for which we can use $l_{u-1}$.

Each node $v\in \LL_i$ ($i > 0$) has at least one neighbour in $\LL_{i-1}$, since the first node in the shortest path from $v$ to $x$ belongs to $\LL_{i-1}$. Consequently, all $\LL_i$ for $i \le u$ are non-empty. Note also that vertices from $\LL_i$ are adjacent only to vertices $\LL_{i-1}$, $\LL_{i}$ and $\LL_{i+1}$ since any edge from $\LL_i$ to $\LL_{j}$ with $j < i - 1$ would imply a shorter path from the node in $\LL_i$ to $x$.

\begin{lemma}
\label{l-u-at-most-2}
A vertex $v\in \LL_i$ can have a degree of at most $k - i$.
\end{lemma}

The vertex $v$ is not adjacent to any vertex in $\LL_j$, where $0 \le j < i - 1$. Since $\LL_j$ are non-empty, there are at least $i - 1$ non-adjacent vertices, so the degree of $v$ is at most $k - 1 - (i - 1) = k - i$.

As a consequence, if $d(l_u)=k-2$, then $u \le 2$.

\begin{lemma}
\label{l-k-2}
If $d(v) = k - 2$ and $v\in L_2$, then $v$ is adjacent to all vertices except $x$.
\end{lemma}

A vertex in $L_2$ is not adjacent to $x$ by definition of $L_2$, and there are no loops, so to have a degree of $k-2$ it must be adjacent to all other vertices.

\begin{lemma}
\label{l-k-1}
If \GG is not a complete graph, then $d(l_u) \le k - 2$
\end{lemma}

For $u > 1$, the lemma follows directly from Lemma~\ref{l-u-at-most-2}, so we only need to prove it for $u=1$. For contrapositive, assume that $d(l_1) = k - 1$. Since $l_1$ has the smallest degree in $\LL_1$, all vertices in $\LL_1$ have a degree of $k - 1$. Furthermore, $x$ has a degree of $k-1$ since all vertices in $\LL_1$ are adjacent to it by definition of $L_1$. Hence, $\GG$ is a complete graph.

The last lemma ensures that the farthest vertex with the lowest degree, $l_u$, fulfills the first condition. It also fulfills the second one: all vertices are connected to $x$ with the shortest paths of lengths at most $u$, which cannot include $l_u$, thus the removal of $l_u$ keeps them connected (at least) via $x$.

We will prove that $l_u$ also fulfills the third condition, except for one special case ($d(l_u) = k - 2$ and $u=2$ and $|\LL_2| = 1$ and $k \ge 5$ and $d(l_1)\le k-2$), in which we choose another suitable vertex. We will consider six different cases, which are (except for the trivial first case) illustrated in Fig.~\ref{f-cases}.

\begin{description}

\item[1. $d(l_u) < k - 2$:] Condition (iii) does not apply.

\item[2. $d(l_u) = k - 2$ and $u=1$:] Since all vertices except $x$ are in $\LL_1$, they are adjacent to $x$ (Fig.~\ref{f-case-1}). $x$ itself is among the neighbours of $l_1$, hence neighbours of $l_1$ are connected through $x$.

\item[3. $d(l_u) = k - 2$ and $u=2$ and $|\LL_2| > 1$:] Since $l_2$ is the vertex with the smallest degree in $\LL_2$, all vertices in $\LL_2$ must have a degree of $k-2$ and are adjacent to all vertices except $x$ by Lemma~\ref{l-k-2} (Fig.~\ref{f-case-2}). The neighbour set of $l_2$ is $\LL_1\cup\LL_2\setminus l_2$. Since $|\LL_2| > 1$, there exists a vertex $v\in\LL_2$ s.t. $v\ne l_2$. $v$ is adjacent to all nodes from $\LL_2\cup\LL_1$, therefore $\LL_1\cup\LL_2\setminus l_2$ is connected.

\item[4. $d(l_u) = k - 2$ and $u=2$ and $|\LL_2| = 1$ and $k = 4$:] $\LL_1$ contains two vertices; both are adjacent to $x$ by definition of $\LL_1$ and to $l_2$ since $d(l_2)=k-2=2$. $l_2$ is not adjacent to $x$ by definition of $\LL_2$. This leaves only two possible graphs, the cycle $C_4$ and a diamond (Fig.~\ref{f-case-3}). For the former, the vertex with the required properties does not exist. For the diamond, $l_u$ fulfills all three conditions.

\item[5. $d(l_u) = k - 2$ and $u=2$ and $|\LL_2| = 1$ and $k \ge 5$ and $d(l_1)=k-1$:] The neighbour set of $l_2$ is the entire $L_1$ (Fig.~\ref{f-case-4}). Since the smallest degree in $L_1$ is $k-1$, $L_1$ is a complete graph and therefore connected.

\item[6. $d(l_u) = k - 2$ and $u=2$ and $|\LL_2| = 1$ and $k \ge 5$ and $d(l_1)\le k-2$:] The graph consists of $\LL_0=\{x\}$, $\LL_2=\{l_2\}$, and of $\LL_1$ with at least 3 vertices since $k \ge 5$ (Fig.~\ref{f-case-6}). All nodes in $\LL_1$ are adjacent to $x$ by definition of $\LL_1$ and to $l_2$ by Lemma~\ref{l-k-2} since we assume $d(l_2)=k-2$.

In this case, $l_u$ does not always fulfil the conditions, so we choose the lowest degree vertex from $\LL_1$, $l_1$. It fulfils the condition (i) by assumptions of this special case. As for condition (ii), the graph $\GG\setminus l_1$ is still connected since all points in $\LL_1$ are adjacent to $x$. Since $|\LL_1| \ge 3$ and $d(l_u) = d(l_2) = k - 2$, vertices $x$ and $l_2$ are connected through the remaining vertices in $\LL_1\setminus l_1$.

Condition (iii) needs to be verified just for the case when $d(l_1) = k - 2$ (Fig.~\ref{f-case-5}). The neighbours of $l_1$ include $x$, $l_2$ and all vertices from $\LL_1$ except one. Since $|\LL_1| \ge 3$, $\LL_1$ must include at least one other neighbour of $l_1$, which thus connects $x$ and $l_u$.
\end{description}

We have covered all possible cases: the degree of $l_u$ cannot exceed $k-2$ due to Lemma~\ref{l-k-1} (assuming the graph is not complete), and when $d(l_u)=k-2$, $u$ cannot exceed 2 due to Lemma~\ref{l-u-at-most-2}.

We have proven that the vertex with the smallest degree in $\LL_u$, $l_u$, fulfills the given conditions in all cases except when $d(l_u) = k - 2$ and $u=2$ and $|\LL_2| = 1$ and $k \ge 5$ and $d(l_1)\le k-2$. In the latter case, the conditions are fulfilled by $l_1$. Complete graphlets and $C_4$ are handled differently.

\subsection{Equation for \texorpdfstring{$C_4$}{C4}}

A cycle on 4 nodes, $C_4$, is treated separately since there is no suitable node $y$ with the required properties. For $C_4$ ($O_8$) we choose one of the nodes adjacent to $x$ for the role of $y$, resulting in 

\begin{equation} 
2o_{8} + 2o_{12} = \sum_{\substack{
x_1,x_2 : \\
x,x_2 \in N(x_1), \\
H[\{x,x_1,x_2\}] \cong G_1}}
\left[ c(x,x_2)-1 \right].
\label{eq-C4}
\end{equation}

Note that this choice violates the third condition that the neighbours of $y$ should induce a connected graph. The equation~\ref{eq-C4} contains a term on the right side that corresponds to the number of common neighbours of node $x$ and some other node $x_2$ at distance 2 from $x$. As further explained in Section~\ref{sec:right}, the algorithm stores precomputed values for all such pairs $x$ and $x_2$, which would require $O(nd^2)$ space and increase the algorithm's space complexity. However, we can still handle this case without consequences for the time and space complexity. We achieve this by reusing $O(n)$ space and recomputing the number of common neighbours every time the algorithm starts a computation of orbit counts for a different node of interest $x$. This optimization is necessary to keep the space requirement at $O(nd)$ for counting four-node graphlets and does not impact the time complexity.

\subsection{System of equations}

In the constructed system of equations, each orbit is related to orbits from graphlets with higher number of edges. This yields a triangular system of equations: we have one equation for every orbit $O$ and these equations include as terms only the orbit $O$ and other orbits belonging to graphlets with a larger number of edges ({\em e.g.}, the orbit 59 in~(\ref{eq-example-59}) is related to orbits 65, 68 and 70). 

The system has $\mathcal{O}_k - 1$ linear equations for $\mathcal{O}_k$ orbit counts. To solve it, one orbit counts must be enumerated directly. The networks that we encounter in practical applications are usually sparse, which makes the complete graphlet (clique) a good candidate. Because of its very few occurrences and its symmetricity, we can efficiently restrict the enumeration. 

Enumerating the orbit in the graph with the largest number of edges also simplifies solving the given triangular system of equations.


\begin{figure}[tb]
\begin{center}
\includegraphics[scale=0.6]{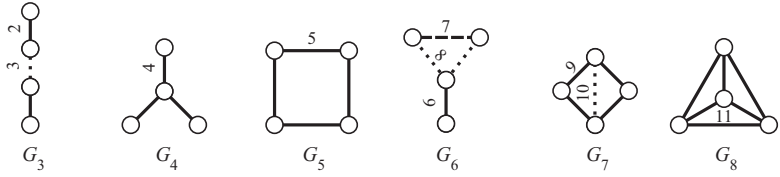}
\caption{Edge orbits in four-node graphlets. Different styles of lines within a graphlet indicate edge orbits.}
\label{four-edge}
\end{center}
\end{figure}

\subsection{Extension to edge orbits}
\label{sec:edge-orbits}
Edge orbits (Figure~\ref{four-edge}) can be defined in a similar way as node orbits. We can use the same approach for setting up the corresponding system of equations. Since the system does not refer to a single $x$ but to an edge ($x_a, x_b$), the selected node $y$ must not coincide with either of these endpoints. We can set $x=x_a$ and show that we can always choose a node $y \neq x_b$ with the required properties.

Since $x_b$ is always in $L_1$, we have to analyze only the cases where we choose $y$ from $L_1$. This happens in cases 1, 2, and 6 from the proof in section \ref{sec:y}. We need to check that there at least two suitable vertices for $y$, so if one of them is $x_b$, the other is chosen as $y$. 

\begin{description}

\item[1. $d(l_u) < k - 2$:] We need to consider only the case when $u=1$. Since $d(l_u) < k - 2$, all vertices in $L_1$ satisfy condition (i). The remaining graph is connected through $x$ (condition (ii)), and condition (iii) does not apply.

\item[2. $d(l_u) = k - 2$ and $u=1$:] Recall that the graph is not complete. Since all vertices in $L_1$ are connected to $x$, there must be at least one pair in $L_1$ that is not connected and thus has a degree of $k - 2$. These two vertices satisfy condition (i). Conditions (ii) and (iii) again hold since all vertices are connected through $x$.

\item[6. $d(l_u) = k - 2$ and $u=2$ and $|\LL_2| = 1$ and $k \ge 5$ and $d(l_1)\le k-2$:] Since the vertex in $L_2$ has a degree of $k-2$, it is connected to all vertices in $L_1$; nodes in $L_1$ are connected to $x$. The nodes in $L_1$ do not induce a complete graph ($d(l_1)\le k-2$), so there must again exist a disconnected pair in $L_1$, which satisfies all conditions like in above case 2.

\end{description}

For $C_4$, one of the nodes in $L_1$ is $x_b$ and the other is $y$.

\section{Algorithm}
\label{sec:right}

Coefficients on the left-hand side of the relations are related to symmetry properties of the graphlets and not to the graph $H$. The terms on the right sides of equations depend on the host graph $H$. Their computation requires enumeration of all graphlets of size $k-1$ and adding up their possible extensions.

The first step is pre-computation and storing of $c(\mathcal{S})$ for all subsets $S$ with up to $k-3$ vertices and for all connected subsets of $k-2$ vertices. These conditions match the criteria for selection of $y$, so the precomputed values $c(\mathcal{S})$ represent the terms in the sum on the right-hand sides of equations.

This is followed by direct enumeration of cliques with $k$ vertices. This enumeration does not have to be extremely fast, but just fast enough not to dominate the time complexity of the entire graphlet counting algorithm. For this purpose we can employ some of the approaches to listing cliques \cite{Bron1973_cliques,Eppstein2010}.

Following this precomputation, the next two steps are repeated for each vertex $x\in V$.

\begin{description}

\item[Computation of sums on the right-hand sides of equations.]

Computation is implemented as enumeration of $(k-1)$-node graphlets touching $x$, as specified by the conditions under the sums. For each graphlet, the terms in the sum consist of the counts $c(\mathcal{S})$ precomputed in the previous step.

For $k \leq 5$, the number of graphlets with $k-1$ nodes is small, so it is feasible to design efficient individual procedures for enumerating them. These procedures involve early pruning of non-viable candidates and completely avoiding any isomorphism testing. Medium-sized graphlets ($k=5$ or $6$) require graphlet recognition of enumerated connected subgraphs, however these patterns can be efficiently distinguished with the use of some trivial invariants such as a \emph{degree sequence}. Enumeration of larger graphlets would benefit from efficient methods for isomorphism testing.

\item[Solving the system of equations.]

The system is triangular, with each equation relating one orbit to those with larger number of edges, Since the count for the highest orbit, which belongs to the clique, is computed by direct enumeration, the system can be solved by decreasing orbit indices.

All orbit counts, coefficients and free terms are integers, thus the computation is numerically stable.
\end{description}

\subsection{Time- and space-complexity}
\label{s-complexity}

We will analyze the worst-case complexity and the expected complexity on random Erd\H{o}s-R\'{e}nyi graphs, followed by empirical verification.

\subsubsection{Worst-case complexity}
\label{sec:worst}

We will evaluate the worst-case time complexity of the algorithm in terms of the number of nodes ($n$) and the maximum degree of a node ($d$) in the host graph. We treat the size of the graphlets, $k$, as a constant. We assume that the graph is stored as a list of adjacent nodes together with a hash table for checking whether two nodes are connected in constant time. The algorithm consists of four steps.

\begin{description}

\item[Precomputation of common neighbours.]

We need to precompute the number of common neighbours of sets of $k-3$ or fewer nodes and of connected sets of $k-2$ nodes to efficiently construct right sides of our equations (Section~\ref{sec:right}). To achieve this we enumerate all subsets of $k-2$ or fewer neighbours for every node. This results in time complexity $O(nd^{k-2})$. Storing the number of common neighbours of sets of at most $k-3$ nodes with the above method requires $O(nd^{k-3})$ space. Because we request that in the case of $k-2$ nodes, these nodes induce a connected subgraph, we can limit their number to the number of $(k-2)$-node induced connected subgraphs, which is also $O(nd^{k-3})$.

\item[Enumeration of cliques.]

We will refer to the time complexity of counting $k$-node cliques in this step as $O(T_k)$. A worst-case time complexity is $O(nd^{k-1})$ and requires constant space. However, this enumeration can be implemented very efficiently in practical applications on sparse networks that contain few cliques.

\item[Enumerating all $(k-1)$-node graphlets and counting their extensions.]

This step computes the right sides of the system of equations. It requires constant space, since the space is reused for each vertex, and runs in $O(nd^{k-2})$ time needed for enumeration of $k-1$-node graphlets.

\item[Solving the system of equations.]

The system of equations is independent of the host graph and requires constant time and space.

\end{description}

Overall, the algorithm has a $O(nd^{k-2} + T_k)$ time complexity while requiring $O(nd^{k-3})$ space. In the worst case, the time complexity is the same as that of a simple exhaustive enumeration method, $O(T_k) = O(nd^{k-1})$. However, the term $T_k$ is much smaller in practice.

\subsubsection{Expected time complexity in random graphs}
\label{sec:expected}

Although the worst-case time complexity of the algorithm is equal to that of brute-force enumeration, the actual performance on real-world networks and on random graphs is much better. We analyzed the expected time complexity on random Erd\H{o}s-R\'{e}nyi graphs with $n$ nodes and edge probability $p$. Throughout this analysis we will assume that $np > 1$, otherwise the graph is likely to have more than one component which can be processed independently.

The precomputation consists of iterating over central nodes, enumerating all sets of $l \leq k-2$ neighbours and incrementing the number of common neighbours of the leaf nodes. The $l$ nodes have to be connected to the central node, which happens with probability $p^l$. The expected time complexity of this step is $O(n \sum_{l=1}^{k-2} n^l p^l)$. Assuming $np > 1$, we can simplify it to $O(n^{k-1}p^{k-2})$.

In the second step, the algorithm enumerates all subgraphs with $k-1$ nodes. It does so incrementally by first enumerating smaller connected subgraphs of size $l$ and extending them to larger connected subgraphs. The expected time complexity is therefore proportional to the expected number of connected subgraphs with $l \leq k-1$ nodes. We need to estimate the probability that a set of $l$ nodes induces a connected subgraph. We can view the process of building every such subgraph by consecutively attaching a new node to at least one of the existing nodes. This of course will overestimate the number of connected subgraphs by some constant because every such subgraph can be built in several different orders of attaching nodes. The probability that an edge exists from some newly added node to at least one of the $i$ existing nodes is $1-(1-p)^i$. The expected number of enumerated subgraphs is therefore $O(\sum_{l=1}^{k-1} n^l \prod_{i=1}^{l-1}(1-(1-p)^i)) = O(\sum_{l=1}^{k-1} n^l p^{l-1})$. Assuming $np > 1$, the expected time complexity is $O(n^{k-1} p^{k-2})$.

The total expected time complexity for setting-up the system of equations in Erd\H{o}s-R\'{e}nyi graphs with $n$ nodes and edge probability $p$ is thus $O(n^{k-1} p^{k-2})$. In practice, we observe graphlets with 4 and 5 nodes. The expected time complexities for these cases are $O(n^3p^2)$ and $O(n^4p^3)$, respectively. 

\subsubsection{Empirical evaluation of time complexity}
\label{sec:empirical}

We evaluated the performance of our algorithm for counting 4- and 5-node graphlets on random Erd\H{o}s-R\'{e}nyi graphs.

We measured the time needed for counting node- and edge-orbits. The running times (Table \ref{tab:node-edge}) are practically the same for counting node-orbits and edge-orbits of both 4- and 5-node graphlets in random graphs with 1\,000 nodes and of increasing density. The size of the graphs ($n=1\,000$) was chosen arbitrarily to put the run times in the range of a couple of seconds. In the remainder of this section we focus on counting node-orbits.

\begin{table*}[tb]
\centering
\caption{Comparison of run times for counting node- and edge-orbits on graphs with
1000 nodes and different densities.}
\vspace{10px}
\label{tab:node-edge}
\begin{tabular}{l|cccc|ccccc}
\multicolumn{1}{c}{}	& \multicolumn{4}{c}{four-node graphlets}	& \multicolumn{5}{c}{five-node graphlets} \\
edges [thousands]		& 50   & 100  & 150  & 200   				& 5    & 10   & 15   & 20   & 25    \\ \hline
node-orbits				& 0.70 & 2.40 & 6.16 & 14.01 				& 0.23 & 1.03 & 2.93 & 6.85 & 13.88 \\
edge-orbits				& 0.69 & 2.33 & 6.12 & 14.21 				& 0.22 & 0.91 & 2.54 & 5.90 & 11.78 \\ \hline
\end{tabular}
\end{table*}

\begin{figure}[bht]
\centering
\includegraphics[bb=3 617 287 789,scale=1.0]{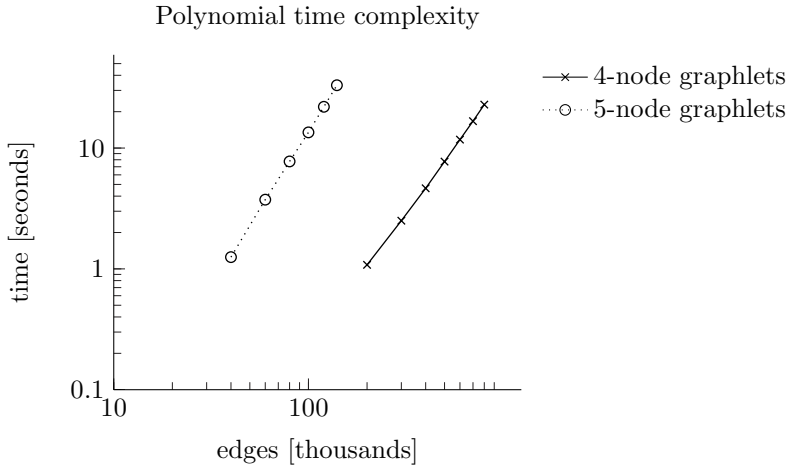}
\caption{Comparison of running times for counting orbits of 4- and 5-node graphlets on sparse graphs with 10000 nodes and varying densities.}
\label{fig:45}
\end{figure}

Second, we compare the running times of counting orbits of 4-node and 5-node graphlets on random graphs with 10\,000 nodes and up to 800\,000 edges. These graphs are sparse as the number of edges represents only about 1.6\% of all possible edges; as such they represent a realistic case from network analysis of large graphs. A logarithmic plot of execution times in Figure \ref{fig:45} shows a polynomial dependence on the size of graphlets. Both plots form a straight line with the steeper one corresponding to counting orbits of 5-node graphlets. 

\begin{figure}[tb]
\centering
\includegraphics[bb=3 617 287 789,scale=1.0]{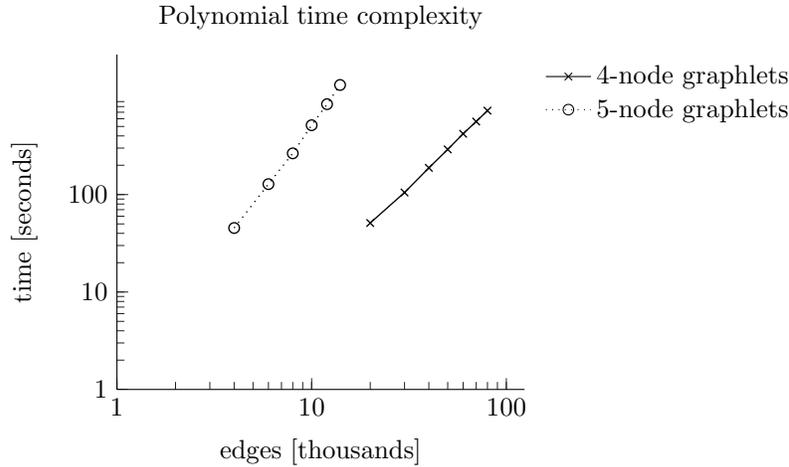}
\caption{Running times with disabled CPU cache.}
\label{fig:cache}
\end{figure}
The slope of the lines should be 2 and 3, respectively, according to the expected time complexities ($O(n^3p^2)$, $O(n^4p^3)$) from Section~\ref{sec:expected}. However, this is clearly not the case in Figure~\ref{fig:45}. Further experiments show that this is the result of CPU cache misses when accessing the precomputed lookup tables. We performed a similar experiment with disabled CPU cache. Because of the slowdown, we decreased the number of nodes to 1\,000 and maintained the ratio of edges to the number of nodes, which is the same as maintaining the average degree of nodes. The measurements with disabled CPU cache in Figure~\ref{fig:cache} line up with the expected slopes of 2 and 3.

\begin{figure}[hbt]
\centering
\includegraphics[bb=3 620 193 789,scale=1.0]{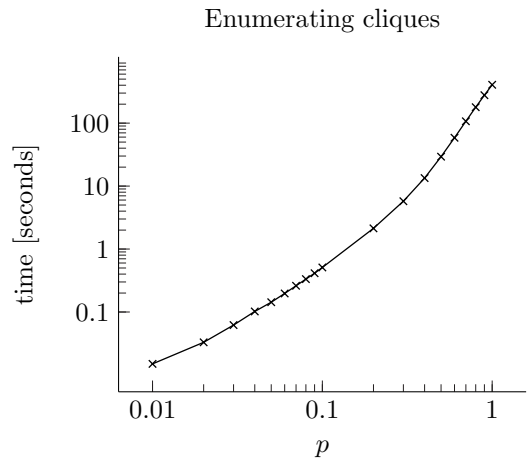}
\caption{Running times on random graphs of increasing density.}
\label{fig:p}
\end{figure}

Finally, we probed for the region in which the enumeration of cliques begins dominating the time complexity. We performed the experiment for counting 4-node graphlets in graphs with 1\,000 nodes and increasing edge probabilities $p$. In Figure \ref{fig:p} the plot follows a straight line up to around $p=0.07$ and another steeper line from $p=0.3$ onwards. This is consistent with the contribution of the step of enumerating cliques. Random sparse graphs contain fewer cliques whose enumeration is efficient and doesn't affect the running time much. However, as the graphs become denser, this becomes the bottleneck of the algorithm.

\section{Final remarks}

The source code of the algorithm in C++, which computes the node and edge orbits for $k=4$ and $k=5$ is available at https://github.com/janezd/orca. The corresponding R package {\tt orca} is also available on CRAN. Parts of this algorithm that have been presented previously are also already included in the GraphCrunch package \cite{Milenkovic2008_graphcrunch}.

\section*{Acknowledgments}

This work has been funded by the Slovenian research agency grants J2-5480 and P2-0209.


\bibliography{main}

\end{document}